\documentclass[aps,10pt,final,
notitlepage, oneside, twocolumn, nobibnotes, nofootinbib,
superscriptaddress, showpacs, centertags, showkeys, assume]{revtex4}

\usepackage{graphicx}
\usepackage{amsmath}

\begin{document}

\title{Sum rules for total cross-sections of hadron  photoproduction
on ground state $1/2^+$ octet baryons}

\author{S.~Dubni\v{c}ka} \affiliation{Inst.\ of Physics,
Slovak Academy of\ Sciences, D\'ubravsk\'a cesta 9, 845 11
Bratislava, Slovak Republic}
\author{A.Z.~Dubni\v ckov\'a}
\affiliation{Dept.\ of Theor.Physics, Comenius University,
842 48 Bratislava, Slovak Republic}
\author{E.~A.~Kuraev} \affiliation{Bogol'ubov\ Laboratory of Theoretical
Physics, JINR, Dubna, Russia}
%\date{\today}

\begin{abstract}
Sum rules  are derived relating Dirac mean square  radii and
anomalous magnetic moments of various couples of the ground state
$1/2^+$ octet  baryons with the convergent integral of the
difference of hadron photoproduction total cross-sections on the
baryons. The chain of inequalities for corresponding total
cross-sections has been found from those sum rules.
\end{abstract}

 \maketitle

Recently, the new sum rule has been derived \cite{Bartos04}
\begin{equation}\label{eq:1}
\frac{1}{3}\langle r_{1p}^2\rangle
-\frac{\kappa_{p}^2}{4m_{p}^2}+\frac{\kappa_{n}^2}{4m_{n}^2}=
\frac{2}{\pi^2\alpha}\int\limits_{\omega_{N}}^\infty
\frac{d\omega}{\omega} \big[\sigma_{tot}^{\gamma {p\to
X}}(\omega)- \sigma_{tot}^{\gamma {n\to X}}(\omega)\big]
\end{equation}
relating Dirac proton mean square radius $\langle r_{1p}^2\rangle$
and anomalous magnetic moments of proton $\kappa_p$ and neutron
$\kappa_n$ to the convergent integral over a difference of the
total proton and neutron photoproduction cross-sections, in which
a mutual cancellation of the rise of the corresponding
cross-sections for $\omega \to \infty$ ($\omega$ is the photon
energy in the laboratory frame), created by the Pomeron exchanges,
was achieved. Using similar ideas the new Cabibbo-Radicati
\cite{Cab66} like sum rules for various suitable couples of the
members of the pseudoscalar meson nonet have been found in Ref.
\cite{dubni06}. In this work, to be fascinated just by the very
precise experimental satisfaction of the sum rule for a difference
of proton and neutron total photoproduction cross-sections
evaluating both sides of (\ref{eq:1}) (for more detail see
\cite{Bartos04}) and getting (1.93$\pm $ 0.18)mb and (1.92$\pm $
0.32)mb, respectively, we extend the method for a derivation of
all possible sum rules for various suitable couples of the members
of the ground state $1/2^+$ octet baryons.

Really, in the first place starting from a very high energy
peripheral electroproduction process on baryon $B$
\begin{equation}
e^-(p_1) + B(p) \to e^-(p_1') + X, \label{eq:2}
\end{equation}
with the produced pure hadronic state $X$  moving closely to the
direction of the initial baryon, one comes (see \cite{Bartos04})
to the Weizs\"acker - Williams like relations, relating the
difference of $\bf q^2$-dependent (the photon transferred
four-momentum $q^2=-\bf q^2$) differential DIS cross-sections with
integral over the difference of the total hadron photoproduction
cross-sections on octet baryons. Then utilizing analytic
properties of parts of forward virtual Compton scattering
amplitudes on octet baryons together with previous result one
comes to the universal (compare with (\ref{eq:1})) octet baryon
sum rule

\begin{eqnarray}\nonumber
&&\frac{1}{3}\big [F_{1B}(0)\langle r_{1B}^2
\rangle-F_{1B'}(0)\langle r_{1B'}^2 \rangle\big ]-
\big [\frac{\kappa_B^2}{4m_B^2}-\frac{\kappa_{B'}^2}{4m^2_{B'}}\big ]=\\
\label{eq:3}
&=&\frac{2}{\pi^2\alpha}\int\limits_{{\omega_B}}^{\infty}
\frac{d\omega} {\omega}\big[\sigma_{tot}^{\gamma B\to X}(\omega)-
\sigma_{tot}^{\gamma B'\to X}(\omega)\big]
\end{eqnarray}
relating Dirac baryon mean square radii $\langle
r_{1B}^2\rangle$,$\langle r_{1B'}^2\rangle$ and baryon anomalous
magnetic moments $\kappa_B , \kappa_{B'}$ to the convergent
integral over a difference of the total baryon photoproduction
cross-sections. As there are 8 different members, $p, n, \Lambda,
\Sigma^+, \Sigma^0, \Sigma^-, \Xi^0, \Xi^-$, of the ground state
$1/2^+$ baryon octet, one can write down 28 different sum rules of
the type (\ref{eq:3}). In order to evaluate the left-hand sides of
them one uses the baryon masses and magnetic moments from Review
of Particle Physics \cite{Rev06} (the $\Sigma^0$ magnetic moment
is determined from the well known relation
$\mu_{\Sigma^+}+\mu_{\Sigma^-}=2 \mu_{\Sigma^0}$) and Dirac baryon
mean square radii $\langle r_{1B}^2\rangle$ are calculated by
means of the relation
\begin{equation}
\langle r_{1B}^2 \rangle =\langle r_{EB}^2
\rangle-\frac{3\kappa_B}{2 m_B^2}, \label{eq:4}
\end{equation}
where the baryon electric mean square radii $\langle r_{EB}^2
\rangle$ are calculated by Kubis and Meissner \cite{Meiss01} to
fourth order in relativistic baryon chiral perturbation theory,
giving predictions for the $\Sigma^-$ charge radius and
$\Lambda-\Sigma^0$ transition moment in excellent agreement with
the available experimental information.

As a result one gets numerically
 { \begin{widetext}
\begin{equation}
\label{eq:5}
\frac{2}{\pi^2\alpha}\int\limits_{\omega_{p}}^{\infty}
\frac{d\omega} {\omega}\big[\sigma_{tot}^{\gamma p\to X}(\omega)-
\sigma_{tot}^{\gamma n\to X}(\omega)\big ]=2.0415
  {\textrm mb},\quad \textrm{
  then in averaged}\quad
\sigma_{tot}^{\gamma p \to X}(\omega)> \sigma_{tot}^{\gamma n\to
X}(\omega)
\end{equation}

\begin{equation}
\label{eq:6}
\frac{2}{\pi^2\alpha}\int\limits_{\omega_{\Sigma^+}}^{\infty}
\frac{d\omega} {\omega}\big[\sigma_{tot}^{\gamma \Sigma^+\to
X}(\omega)- \sigma_{tot}^{\gamma \Sigma^0\to X}(\omega)\big
]=2.0825
  {\textrm mb},\quad \textrm{
  then in averaged}\quad
\sigma_{tot}^{\gamma \Sigma^+\to X}(\omega)> \sigma_{tot}^{\gamma
\Sigma^0\to X}(\omega)
\end{equation}

\begin{equation}
\label{eq:7}
\frac{2}{\pi^2\alpha}\int\limits_{\omega_{\Sigma^+}}^{\infty}
\frac{d\omega} {\omega}\big[\sigma_{tot}^{\gamma \Sigma^+\to
X}(\omega)- \sigma_{tot}^{\gamma \Sigma^-\to X}(\omega) \big
]=4.2654  {\textrm mb}, \quad \textrm{
  then in averaged}\quad \sigma_{tot}^{\gamma \Sigma^+\to
X}(\omega)> \sigma_{tot}^{\gamma \Sigma^-\to X}(\omega)
\end{equation}

\begin{equation}
\label{eq:8}
\frac{2}{\pi^2\alpha}\int\limits_{\omega_{\Sigma^0}}^{\infty}
\frac{d\omega} {\omega}\big[\sigma_{tot}^{\gamma \Sigma^0\to
X}(\omega)- \sigma_{tot}^{\gamma \Sigma^-\to X}(\omega)\big ]=
2.1829  {\textrm mb}, \quad \textrm{
  then in averaged}\quad \sigma_{tot}^{\gamma \Sigma^0\to
X}(\omega)> \sigma_{tot}^{\gamma \Sigma^-\to X}(\omega)
\end{equation}

\begin{equation}
\label{eq:9}
\frac{2}{\pi^2\alpha}\int\limits_{\omega_{\Xi^0}}^{\infty}
\frac{d\omega} {\omega}\big[\sigma_{tot}^{\gamma \Xi^0\to
X}(\omega)- \sigma_{tot}^{\gamma \Xi^-\to X}(\omega)\big ]=1.5921
    {\textrm mb}, \quad \textrm{then in averaged}\quad \sigma_{tot}^{\gamma
\Xi^0\to X}(\omega)> \sigma_{tot}^{\gamma \Xi^-\to X}(\omega)
\end{equation}

\begin{equation}
\label{eq:10}
\frac{2}{\pi^2\alpha}\int\limits_{\omega_{p}}^{\infty}
\frac{d\omega} {\omega}\big[\sigma_{tot}^{\gamma p\to X}(\omega)-
\sigma_{tot}^{\gamma \Lambda^0\to X}(\omega)\big ]=1.6673 {\textrm
mb},\quad \textrm{then in averaged}\quad \sigma_{tot}^{\gamma p\to
X}(\omega)> \sigma_{tot}^{\gamma \Lambda^0\to X}(\omega)
\end{equation}

\begin{equation}
\label{eq:11}
\frac{2}{\pi^2\alpha}\int\limits_{\omega_{p}}^{\infty}
\frac{d\omega} {\omega}\big[\sigma_{tot}^{\gamma p \to X}(\omega)-
\sigma_{tot}^{\gamma \Sigma^+\to X}(\omega)\big ]=-0.4158 {\textrm
mb},\quad \textrm{then in averaged}\quad \sigma_{tot}^{\gamma p
\to X}(\omega)<\sigma_{tot}^{\gamma \Sigma^+\to X}(\omega)
\end{equation}

\begin{equation}
\label{eq:12}
\frac{2}{\pi^2\alpha}\int\limits_{\omega_{p}}^{\infty}
\frac{d\omega} {\omega}\big[\sigma_{tot}^{\gamma p\to X}(\omega)-
\sigma_{tot}^{\gamma \Sigma^0\to X}(\omega)\big ]=1.6667  {\textrm
mb},\quad \textrm{then in averaged}\quad \sigma_{tot}^{\gamma p\to
X}(\omega)> \sigma_{tot}^{\gamma \Sigma^0\to X}(\omega)
\end{equation}

\begin{equation}
\label{eq:13}
\frac{2}{\pi^2\alpha}\int\limits_{\omega_{p}}^{\infty}
\frac{d\omega} {\omega}\big[\sigma_{tot}^{\gamma p \to X}(\omega)-
\sigma_{tot}^{\gamma \Sigma^-\to X}(\omega)\big ]= 3.8496 {\textrm
mb},\quad \textrm{then in averaged}\quad \sigma_{tot}^{\gamma p
\to X}(\omega)> \sigma_{tot}^{\gamma \Sigma^-\to X}(\omega)
\end{equation}

\begin{equation}
\label{eq:14}
\frac{2}{\pi^2\alpha}\int\limits_{\omega_{p}}^{\infty}
\frac{d\omega} {\omega}\big[\sigma_{tot}^{\gamma p\to X}(\omega)-
\sigma_{tot}^{\gamma \Xi^0\to X}(\omega)\big ]=1.7259  {\textrm
mb},\quad \textrm{then in averaged}\quad \sigma_{tot}^{\gamma p\to
X}(\omega)> \sigma_{tot}^{\gamma \Xi^0\to X}(\omega)
\end{equation}

\begin{equation}
\label{eq:15}
\frac{2}{\pi^2\alpha}\int\limits_{\omega_{p}}^{\infty}
\frac{d\omega} {\omega}\big[\sigma_{tot}^{\gamma p \to X}(\omega)-
\sigma_{tot}^{\gamma \Xi^-\to X}(\omega)\big ]=3.3180   {\textrm
mb},\quad \textrm{then in averaged}\quad \sigma_{tot}^{\gamma p
\to X}(\omega)> \sigma_{tot}^{\gamma \Xi^-\to X}(\omega)
\end{equation}

\begin{equation}
\label{eq:16}
\frac{2}{\pi^2\alpha}\int\limits_{\omega_{n}}^{\infty}
\frac{d\omega} {\omega}\big[\sigma_{tot}^{\gamma n\to X}(\omega)-
\sigma_{tot}^{\gamma \Lambda^0\to X}(\omega)\big ]=-0.3260
{\textrm mb},\quad \textrm{then in averaged}\quad
\sigma_{tot}^{\gamma n\to X}(\omega)< \sigma_{tot}^{\gamma
\Lambda^0\to X}(\omega)
\end{equation}

\begin{equation}
\label{eq:17}
\frac{2}{\pi^2\alpha}\int\limits_{\omega_{n}}^{\infty}
\frac{d\omega} {\omega}\big[\sigma_{tot}^{\gamma n\to X}(\omega)-
\sigma_{tot}^{\gamma \Sigma^+\to X}(\omega)\big ]=-2.4573 {\textrm
mb},\quad \textrm{then in averaged}\quad \sigma_{tot}^{\gamma n\to
X}(\omega)< \sigma_{tot}^{\gamma \Sigma^+\to X}(\omega)
\end{equation}

\begin{equation}
\label{eq:18}
\frac{2}{\pi^2\alpha}\int\limits_{\omega_{n}}^{\infty}
\frac{d\omega} {\omega}\big[\sigma_{tot}^{\gamma n\to X}(\omega)-
\sigma_{tot}^{\gamma \Sigma^0\to X}(\omega)\big ]=-0.3747 {\textrm
mb},\quad \textrm{then in averaged}\quad \sigma_{tot}^{\gamma n\to
X}(\omega)< \sigma_{tot}^{\gamma \Sigma^0\to X}(\omega)
\end{equation}

\begin{equation}
\label{eq:19}
\frac{2}{\pi^2\alpha}\int\limits_{\omega_{n}}^{\infty}
\frac{d\omega} {\omega}\big[\sigma_{tot}^{\gamma n\to X}(\omega)-
\sigma_{tot}^{\gamma \Sigma^-\to X}(\omega)\big ]= 1.8082 {\textrm
mb},\quad \textrm{then in averaged}\quad \sigma_{tot}^{\gamma n\to
X}(\omega)> \sigma_{tot}^{\gamma \Sigma^-\to X}(\omega)
\end{equation}

\begin{equation}
\label{eq:20}
\frac{2}{\pi^2\alpha}\int\limits_{\omega_{n}}^{\infty}
\frac{d\omega} {\omega}\big[\sigma_{tot}^{\gamma n\to X}(\omega)-
\sigma_{tot}^{\gamma \Xi^0\to X}(\omega)\big ]= -0.3156  {\textrm
mb},\quad \textrm{then in averaged}\quad \sigma_{tot}^{\gamma n\to
X}(\omega)< \sigma_{tot}^{\gamma \Xi^0\to X}(\omega)
\end{equation}

\begin{equation}
\label{eq:21}
\frac{2}{\pi^2\alpha}\int\limits_{\omega_{n}}^{\infty}
\frac{d\omega} {\omega}\big[\sigma_{tot}^{\gamma n\to X}(\omega)-
\sigma_{tot}^{\gamma \Xi^-\to X}(\omega)\big ]=1.2766  {\textrm
mb},\quad \textrm{then in averaged}\quad \sigma_{tot}^{\gamma n\to
X}(\omega)> \sigma_{tot}^{\gamma \Xi^-\to X}(\omega)
\end{equation}

\begin{equation}
\label{eq:22}
\frac{2}{\pi^2\alpha}\int\limits_{\omega_{\Lambda^0}}^{\infty}
\frac{d\omega} {\omega}\big[\sigma_{tot}^{\gamma \Lambda^0\to
X}(\omega)- \sigma_{tot}^{\gamma \Sigma^+\to X}(\omega)\big ]=
-2.0831 {\textrm mb},\quad \textrm{then in averaged}\quad
\sigma_{tot}^{\gamma \Lambda^0\to X}(\omega)< \sigma_{tot}^{\gamma
\Sigma^+\to X}(\omega)
\end{equation}

\begin{equation}
\label{eq:23}
\frac{2}{\pi^2\alpha}\int\limits_{\omega_{\Lambda^0}}^{\infty}
\frac{d\omega} {\omega}\big[\sigma_{tot}^{\gamma \Lambda^0\to
X}(\omega)- \sigma_{tot}^{\gamma \Sigma^0\to X}(\omega)\big ]=
-0.0006   {\textrm mb},\quad \textrm{then in averaged}\quad
\sigma_{tot}^{\gamma \Lambda^0\to X}(\omega)\approx
\sigma_{tot}^{\gamma \Sigma^0\to X}(\omega)
\end{equation}

\begin{equation}
\label{eq:24}
\frac{2}{\pi^2\alpha}\int\limits_{\omega_{\Lambda^0}}^{\infty}
\frac{d\omega} {\omega}\big[\sigma_{tot}^{\gamma \Lambda^0\to
X}(\omega)- \sigma_{tot}^{\gamma \Sigma^-\to X}(\omega)\big
]=2.1823
  {\textrm{mb}},\quad \textrm{then in averaged}\quad \sigma_{tot}^{\gamma
\Lambda^0\to X}(\omega)> \sigma_{tot}^{\gamma \Sigma^-\to X}(\omega)
\end{equation}

\begin{equation}
\label{eq:25}
\frac{2}{\pi^2\alpha}\int\limits_{\omega_{\Lambda^0}}^{\infty}
\frac{d\omega} {\omega}\big[\sigma_{tot}^{\gamma \Lambda^0\to
X}(\omega)- \sigma_{tot}^{\gamma \Xi^0\to X}(\omega)\big ]=0.0586
{\textrm{mb}},\quad \textrm{then in averaged}\quad
\sigma_{tot}^{\gamma \Lambda^0\to X}(\omega)> \sigma_{tot}^{\gamma
\Xi^0\to X}(\omega)
\end{equation}

\begin{equation}
\label{eq:26}
\frac{2}{\pi^2\alpha}\int\limits_{\omega_{\Lambda^0}}^{\infty}
\frac{d\omega} {\omega}\big[\sigma_{tot}^{\gamma \Lambda^0\to
X}(\omega)- \sigma_{tot}^{\gamma \Xi^-\to X}(\omega)\big ]=2.1823
{\textrm{mb}} ,\quad \textrm{then in averaged}\quad
\sigma_{tot}^{\gamma \Lambda^0\to X}(\omega)> \sigma_{tot}^{\gamma
\Xi^-\to X}(\omega)
\end{equation}

\begin{equation}
\label{eq:27}
\frac{2}{\pi^2\alpha}\int\limits_{\omega_{\Sigma^+}}^{\infty}
\frac{d\omega} {\omega}\big[\sigma_{tot}^{\gamma \Sigma^+\to
X}(\omega)- \sigma_{tot}^{\gamma \Xi^0\to X}(\omega)\big ]=2.1417
\textrm{mb},\quad \textrm{then in averaged}\quad
\sigma_{tot}^{\gamma \Sigma^+\to X}(\omega)> \sigma_{tot}^{\gamma
\Xi^0\to X}(\omega)
\end{equation}

\begin{equation}
\label{eq:28}
\frac{2}{\pi^2\alpha}\int\limits_{\omega_{\Sigma^+}}^{\infty}
\frac{d\omega} {\omega}\big[\sigma_{tot}^{\gamma \Sigma^+\to
X}(\omega)- \sigma_{tot}^{\gamma \Xi^-\to X}(\omega)\big ]=3.7338
\textrm{mb},\quad \textrm{then in averaged}\quad
\sigma_{tot}^{\gamma \Sigma^+\to X}(\omega)> \sigma_{tot}^{\gamma
\Xi^-\to X}(\omega)
\end{equation}

\begin{equation}
\label{eq:29}
\frac{2}{\pi^2\alpha}\int\limits_{\omega_{\Sigma^0}}^{\infty}
\frac{d\omega} {\omega}\big[\sigma_{tot}^{\gamma \Sigma^0\to
X}(\omega)- \sigma_{tot}^{\gamma \Xi^0\to X}(\omega)\big ]=0.1168
\textrm{mb},\quad \textrm{then in averaged}\quad
\sigma_{tot}^{\gamma \Sigma^0\to X}(\omega)> \sigma_{tot}^{\gamma
\Xi^0\to X}(\omega)
\end{equation}

\begin{equation}
\label{eq:30}
\frac{2}{\pi^2\alpha}\int\limits_{\omega_{\Sigma^0}}^{\infty}
\frac{d\omega} {\omega}\big[\sigma_{tot}^{\gamma \Sigma^0\to
X}(\omega)- \sigma_{tot}^{\gamma \Xi^-\to X}(\omega)\big ]=1.5732
\textrm{mb},\quad \textrm{then in averaged}\quad
\sigma_{tot}^{\gamma \Sigma^0\to X}(\omega)> \sigma_{tot}^{\gamma
\Xi^\to X}(\omega)
\end{equation}

\begin{equation}
\label{eq:31}
\frac{2}{\pi^2\alpha}\int\limits_{\omega_{\Sigma^-}}^{\infty}
\frac{d\omega} {\omega}\big[\sigma_{tot}^{\gamma \Sigma^-\to
X}(\omega)- \sigma_{tot}^{\gamma \Xi^0\to X}(\omega)\big ]=-2.1238
\textrm{mb},\quad \textrm{then in averaged}\quad
\sigma_{tot}^{\gamma \Sigma^-\to X}(\omega)< \sigma_{tot}^{\gamma
\Xi^0\to X}(\omega)
\end{equation}

\begin{equation}
\label{eq:32}
\frac{2}{\pi^2\alpha}\int\limits_{\omega_{\Sigma^-}}^{\infty}
\frac{d\omega} {\omega}\big[\sigma_{tot}^{\gamma \Sigma^-\to
X}(\omega)- \sigma_{tot}^{\gamma \Xi^-\to X}(\omega)\big ]=-0.5316
\textrm{mb},\quad \textrm{then in averaged}\quad
\sigma_{tot}^{\gamma \Sigma^-\to X}(\omega)< \sigma_{tot}^{\gamma
\Xi^-\to X}(\omega),
\end{equation}
\end{widetext}}

from where the following chain of inequalities {\begin{widetext}
\begin{equation}
\label{eq:33}
 \sigma_{tot}^{\gamma \Sigma^+\to X}(\omega)>
\sigma_{tot}^{\gamma p\to X}(\omega)> \sigma_{tot}^{\gamma
\Lambda^0\to X}(\omega) \approx   \sigma_{tot}^{\gamma \Sigma^0\to
X}(\omega)> \sigma_{tot}^{\gamma \Xi^0\to X}(\omega)>
 \sigma_{tot}^{\gamma n\to X}(\omega)>
\sigma_{tot}^{\gamma \Xi^-\to X}(\omega)>\sigma_{tot}^{\gamma
\Sigma^-\to X}(\omega)
\end{equation}
\end{widetext}}
for total cross-sections of hadron photoproduction on ground state
$1/2^+$ octet baryons is found.

Experimental tests of the derived sum rules, as well as
inequalities (\ref{eq:33}), could be practically carried out
provided there exist data on total hadron photoproduction
cross-sections on hyperons as a function of energy which, however,
 are missing up to present time. Nevertheless, the idea of
intensive photon beams generated by the electron beams of linear
$e^+e^-$ colliders by using the process of the backward Compton
scattering of laser light off the high energy electrons
\cite{Ginzb81} is encouraging and one expects that a measurements
of the total hadron photoproduction cross-sections on hyperons
could be practically achievable in future.

\bigskip

The work was partly supported by Slovak Grant Agency for Sciences
VEGA, Grant No. 2/4099/26 (S.D. and A.Z.D).  A.Z. Dubni\v ckov\'a
would like to thank University di Trieste for warm hospitality at
the early stage of this work and Professor N. Paver for numerous
doscussions.

\end{document}